\documentstyle[aps,epsf]{revtex}
\def \be   {\begin{equation}}
\def \ee   {\end{equation}}
\def \l {\label}
\begin{document}
\input epsf
\baselineskip=25pt
\title{Discrete interactions and the Pioneer anomalous acceleration: Alternative II}
\author{Manoelito M de Souza}
\address{{Departamento de
F\'{\i}sica - Universidade Federal do Esp\'{\i}rito Santo\\29065.900 -Vit\'oria-ES-Brazil}\thanks{ E-mail: manoelit@verao.cce.ufes.br}}
\date{\today}
\maketitle
\begin{abstract} The dominant contributions from a discrete gravitational interaction produce the standard potential as an effective continuous field. The sub-dominant contributions are, in a first approximation, linear on $n$, the accumulated  number of (discrete) interaction events along the test-body trajectory. For a nearly radial trajectory $n$ is proportional to the transversed distance and its effects may have been observed as the Pioneer anomalous constant radial acceleration, which cannot be observed on the nearly circular planetary orbits. Here we give calculation details of the alternative II, discussed in gr-qc/0106046.
\end{abstract}
\begin{center}
\end{center}

The constant anomalous radial acceleration observed \cite{gr-qc/0104064,Pioneer} in the Pioneer 10 and 11 spacecrafts defies explanation. Discrete interactions can explain it; here we give calculation details of the alternative II, discussed in \cite{gr-qc/0106046}. 
We consider a non-relativistic radial motion with a discrete interaction whose effective description is an acceleration field that decreases with the inverse of the squared distance. The much simpler case of a radial motion with a non-relativistic, axially symmetric, interaction (where the effective acceleration is inversely proportional to the distance) has been considered in \cite{hep-th/0103218}.

Assuming that the time interval between two consecutive interactions is the (statistical average) two-way flying time between two interacting elementary point objects (components), then 
\be
\l{dti}
\Delta t_{i}=\alpha r_{i},
\ee
where $r_{i}$ is their space separation at the $i^{th}$ interaction. This requires that the change in speed at each interaction event be given by 
\be
\l{dv}
\Delta v_{i}\equiv\frac{K}{r_{i}},
\ee
in order to reproduce the observed effective Newtonian acceleration. See the reference \cite{hep-th/9610145} for a physical interpretation.
$\alpha$ and $K$ are constants to be determined later. Rigourously, instead of a change of velocity , we should consider a change of four momentum, representing the energy-momentum carried by a quantum of (discrete) interaction, but weak fields and low velocities justify this non-relativistic simplification. The rest frame of a central source, a mass $M$ in the case of gravity, is then assumed. Clearly this is an effective description which is denounced by the singularity (infinity) at the origin. It does not fit the discrete-field filosophy which implies finite interactions but, like the alternative I, both discussed in \cite{gr-qc/0106046}, it reproduces the standard continuous gravity and the observed anomalous Pioneer acceleration. 

For initial conditions taken as $$r(t_{0})= r_{0};\qquad v(t_{o})=v_{0},$$ the next interaction will occur at 
\be
\l{Dt0}
t_{1}= t_{0}+\Delta t_{0}\equiv t_{0}+\alpha r_{0},
\ee
neglecting relativistic corrections.   
 Then, 
\be
v(t_{1})\equiv v_{1}=v_{0}-\frac{K}{r_{1}},
\ee
and
\be
\l{r1}
r(t_{1})\equiv r_{1}= r_{0}+v_{0}\Delta t_{0}=(1+\alpha v_{0})r_{0},
\ee
as there is free propagation between any two consecutive interaction events.
Therefore, in the $n^{th}$ interaction 
\be
\l{rn}
r_{n}=r_{n-1}+v_{n-1}\Delta t_{n-1}=(1+\alpha v_{n-1})r_{n-1}=r_{0}\amalg_{i=0}^{n-1}(1+\alpha v_{i}),
\ee
 with 
\be
\l{vj}
v_{j}=v_{0}-\sum_{i=1}^{j}\frac{K}{r_{i}}.
\ee
Then
\be
x_{n}\equiv\frac{r_{n}}{r_{0}}= 1+\alpha\sum_{i_{1}=0}^{n-1}v_{i_{1}}+\alpha^2\sum_{i_{1}=0}^{n-1}\sum_{i_{2}=i_{1}+1}^{n-1}v_{i_{1}}v_{i_{2}}\dots+\alpha^{n-1}(\sum_{i_{1}=0}^{n-1}\sum_{i_{2}=i_{1}+1}^{n-1}\dots \sum_{i_{n-1}=i_{n-2}+1}^{n-1})v_{i_{1}}v_{i_{2}}\dots v_{n-1},
\ee
or
\be
\l{rn0}
\frac{r_{n}}{r_{0}}= 1+\alpha\sum_{i_{1}=0}^{n-1}v_{i_{1}}{\Big\{}1+\alpha\sum_{i_{2}=i_{1}+1}^{n-1}v_{i_{2}}{\Big\{}+\dots+\sum_{i_{n-1}=i_{n-2}+1}^{n-1}v_{n-1}{\Big\}}\dots{\Big\}},
\ee
in a more compact notation.
This is a finite series of $n+1$ terms. Our objective is, of course, writing $r_{n}$ and $v_{n}$ in terms of the initial conditions $r_{0}$ and $v_{0}.$ It is, however, convenient to keep the expressions of $v_{i}$ in the intermediary step, in terms of $r_{n}$, $n>i$, which is to be replaced, at the end, by its final expression in terms of the initial conditions. From the Eq. (\ref{rn}) we may write
\be \l{rnj}
\frac{r_{n}}{r_{j}}=\amalg_{i=j}^{n-1}(1+\alpha v_{i})=1+\alpha\sum_{i_{1}=j}^{n-1}v_{i_{1}}{\Big\{}1+\alpha\sum_{i_{2}=i_{1}+1}^{n-1}v_{i_{2}}{\Big\{}+\dots+\sum_{i_{n-1}=i_{n-2}+1}^{n-1}v_{n-1}{\Big\}}\dots{\Big\}},
\ee
and so,  from the Eq. (\ref{vj})
\be
\l{vnj}
v_{j}=v_{0}-\frac{K}{r_{n}}\sum_{i=1}^{j}\frac{r_{n}}{r_{i}}=v_{0}-\gamma_{n}{\Big\{}1+\alpha\sum_{i_{1}=j}^{n-1}v_{i_{1}}{\Big\{}1+\alpha\sum_{i_{2}=i_{1}+1}^{n-1}v_{i_{2}}{\Big\{}+\dots+\sum_{i_{n-1}=i_{n-2}+1}^{n-1}v_{n-1}{\Big\}}\dots{\Big\}}{\Big\}},
\ee
with $\gamma_{n}=\frac{K}{r_{n}}$. The Eqs. (\ref{dti},\ref{dv}) replace the diferential equation of the contiuous fields whereas the $(n+1)$-term finite series  (\ref{rnj},\ref{vnj}) replace their respective continuous solutions. As $n$ usually is a huge integer the successive sums may become quite involved  and so it is convenient to adopt a systematic approach using 
\be
 {n\choose 0}=\cases{0,&if $n<k$;\cr
\cr
_1,&if $k=0$;\cr
\cr
\frac{n!}{(n-k)!k!},&if $n\ge k\ge0$,\cr}
\ee
 with
\be
\l{eeq}
\sum_{i=0}^{n-1}{i\choose k}={n\choose k+1},
\ee
$$\sum_{i=1}^{n}\equiv\sum_{i=0}^{n-1}+\delta_{i}^{n}-\delta_{i}^{0},$$
$$\sum_{i=0}^{n-1}i{i\choose k}=(k+1){n\choose k+2}+k{n\choose k+1},$$
$$\sum_{i=0}^{n-1}i^K{i\choose k'}=\frac{k+k'}{k'!}{n\choose k+k'+1}+{\hbox{smaller order terms}},$$

and other results derived from these \cite{gr-qc/0106046} for a systematic writing of each term of this series in terms of combinatorials ${n\choose k}$. 
Let us introduce the following expansions
\be
\l{vs}
v_{n}=\sum_{s=0}^{s=n}\alpha^s V^{(s)}_{n},
\ee
\be
\l{xs}
x_{n}\equiv\frac{r_{n}}{r_{0}}=\sum_{s=0}^{s=n}\alpha^s X^{(s)}_{n},
\ee
and 
\be
\l{vnjs}
v_{j,n}=\sum_{s=0}^{s=n}\alpha^s V^{(s)}_{j,n},
\ee
where $v_{j,n}$ represents $v_{j}$ expressed in terms of $r_{n}$ and not of $r_{j}$, $v_{n,n}=v_{n}$, $V^{(s)}_{n,n}=v_{n}\delta^{s}_{0}$.
Then we have, from Eqs. (\ref{vnj},\ref{vs},\ref{vnjs}),  
\be
\l{Vjn0}
V_{j,n}^{(0)}=v_{0}-\gamma_{n}\sum_{i=1}^{j}1=v_{0}-\gamma_{n}{j\choose 1},
\ee 
and for $n>k\ge1$,
\be
\l{Vnjs}
V_{j,n}^{(k)}= -\gamma_{n}\sum_{i=1}^{j}\sum_{m_{1}=i}^{n-1}{\Big\{}V_{m_{1},n}^{(k-1)}+\sum_{m_{2}=m_{1}+1}^{n-1}{\big\{}(V_{m_{1},n}V_{m_{2},n})^{(k-2)}+\dots+\sum_{m_{k-1}=m_{k-2}+1}^{n-1}(V_{m_{1},n}V_{m_{2},n}\dots V_{m_{k},n})^{(0)}{\big\}}\dots{\Big\}},
\ee
where 
\be
\l{VV}
(V_{m_{1},n}V_{m_{2},n})^{(k)}\equiv \sum_{s=0}^{s=k}V_{m_{1},n}^{(s)}V_{m_{2},n}^{(k-s)},
\ee
\be
\l{VVV}
(V_{m_{1},n}V_{m_{2},n}V_{m_{3},n})^{(k)}\equiv \sum_{s=0}^{s=k}V_{m_{1},n}^{(s)}(V_{m_{2},n}V_{m_{3},n})^{(k-s)},
\ee
and so on. For example
$$
V_{j,n}^{(1)}=-\gamma_{n}\sum_{i=1}^{j}\sum_{m_{1}=i}^{n-1}V_{m_{1},n}^{(0)}=-\gamma_{n}\sum_{i=1}^{j}\sum_{m_{1}=i}^{n-1}v_{0}-\gamma_{n}{j\choose 1}=$$
\be
=-\gamma_{n}{\Big\{}v_{0}{\Big[}{n\choose 1}{j\choose 1}-{j\choose 2}-{j\choose 1}{\Big]}-\gamma_{n}{\Big[}{n\choose 2}{j\choose 1}-{j\choose 3}-{j\choose 2}{\Big]}{\Big\}},
\ee

$$
V_{j,n}^{(2)}=-\gamma_{n}\sum_{i=1}^{j}\sum_{m_{1}=i}^{n-1}{\Big\{}V_{m_{1},n}^{(1)}+\sum_{m_{2}=m_{1}+1}^{n-1}{\big\{}(V_{m_{1},n}V_{m_{2},n})^{(0)}{\big\}}{\Big\}}=
$$
\be
=-\gamma_{n}{\Big\{}v_{0}[{n\choose 1}{j\choose 1}-{j\choose 2}-{j\choose 1}]-\gamma_{n}[{n\choose 2}{j\choose 1}-{j\choose 3}-{j\choose 2}]{\Big\}},
\ee
$$V_{j,n}^{(3)}= -\gamma_{n}\sum_{i=1}^{j}\sum_{m_{1}=i}^{n-1}{\Big\{}V_{m_{1},n}^{(2)}+\sum_{m_{2}=m_{1}+1}^{n-1}{\big\{}(V_{m_{1},n}V_{m_{2},n})^{(1)}+\sum_{m_{3}=m_{2}+1}^{n-1}(V_{m_{1},n}V_{m_{2},n}V_{m_{3},n})^{(0)}{\big\}}{\big\}}{\Big\}},$$
and so on. 
Also, from Eqs. (\ref{rn0},\ref{xs},\ref{vnjs}) we have
$$
X_{n}^{(k)}=\sum_{m_{1}=0}^{n-1}{\Big\{}V_{m_{1},n}^{(k-1)}+\sum_{m_{2}=m_{1}+1}^{n-1}{\Big\{}(V_{m_{1},n}V_{m_{2},n})^{(k-2)}+\sum_{m_{3}=m_{2}+1}^{n-1}{\Big\{}(V_{m_{1},n}V_{m_{2},n}V_{m_{3},n})^{(k-3)}+\dots 
$$
\be
\l{Xn}
\dots+\sum_{m_{k-1}=m_{k-2}+1}^{n-1}(V_{m_{1},n}V_{m_{2}}\dots V_{m_{k},n})^{0}{\Big\}}\dots{\Big\}}.
\ee
For example,

\be
\l{Xn0}
X_{n}^{(0)}=1,
\ee

\be
\l{Xn1}
X_{n}^{(1)}=\sum_{m_{1}=0}^{n-1}V_{m_{1},n}^{(0)}=\sum_{m_{1}=0}^{n-1}(v_{0}-\gamma_{n}{i\choose 1})=v_{0}{n\choose 1}-\gamma_{n}{n\choose 2},
\ee
\be
X_{n}^{(2)}=\sum_{m_{1}=0}^{n-1}{\Big\{}V_{m_{1},n}^{(1)}+\sum_{m_{2}=m_{1}+1}^{n-1}{\Big\{}(V_{m_{1},n}V_{m_{2},n})^{(0)}{\Big\}}{\Big\}}=v^{2}_{0}{n\choose 2}-\frac{v_{0}\gamma_{n}}{6}{\Big\{}5n^{3}-9n^{2}+4n{\Big\}}+\frac{\gamma^{2}_{n}}{6}2n^{4}+5n^{3}+4n^{2}-n{\Big\}},
\ee

$$
X_{n}^{(3)}=\sum_{m_{1}=0}^{n-1}{\Big\{}V_{m_{1},n}^{(2)}+\sum_{m_{2}=m_{1}+1}^{n-1}{\Big\{}(V_{m_{1},n}V_{m_{2},n})^{(1)}+{\Big\{}\sum_{m_{3}=m_{2}+1}^{n-1}(V_{m_{1},n}V_{m_{2},n}V_{m_{3},n})^{(0)}{\Big\}}{\Big\}}{\Big\}},
$$ and so on.

Eqs. (\ref{Vnjs},\ref{Xn}) are recursion relations that allow the complete determination of both finite series (\ref{vs}) and (\ref{xs}). So, considering that generally $n$ is a huge number, $v_{n}$ and $r_{n}$ can, at least in principle, exactly be determined from the knowledge of $v_{0}$ and $r_{0}$. On the other hand, for a huge $n$, we can consider only the dominant contribution from each term in each series. Then we have
\be
\l{svn}
v_{n}\simeq{\Big \{}v_{0}-n\gamma_{n}{\Big \}}-\alpha{\Big \{}n^2\gamma_{n}(\frac{v_{0}}{2}-\frac{n\gamma_{n}}{3}){\Big \}}-\alpha^2{\Big \{}\frac{n^3\gamma_{n}}{3}(\frac{v_{0}^2}{2}-\frac{5}{4}v_{0}n\gamma_{n}+\frac{4}{5}n^2\gamma_{n}){\Big \}}-\dots
\ee
and
\be
\l{sxn}
x_{n}\simeq1+\alpha{\Big \{}n(v_{0}-\frac{n\gamma_{n}}{2}){\Big \}}+\alpha^2{\Big \{}\frac{n}{2}(v_{0}^2-\frac{6}{3}nv_{0}\gamma_{n})+\frac{2n^2\gamma_{n}^2}{3}){\Big \}}+\dots
\ee
We can use the terms in the series (\ref{svn}), according to Eq. (\ref{vs}), for writing the series (\ref{sxn}) as
\be
\l{xsn1}
\frac{r_{n}}{r_{0}}=1+\alpha_{II}{\Big \{}\frac{v_{0}^2-V_{n}^{(0)^{2}}}{2\gamma_{n}}{\Big \}} -\alpha_{II}^2{\Big \{}\frac{2v_{0}V_{n}^{(1)}}{2\gamma_{n}}{\Big \}}-\alpha_{II}^3{\Big\{} 
\frac{V_{n}^{(1)^{2}}+2V_{n}^{(0)}V_{n}^{(2)}}{2\gamma_{n}}{\Big \}}-\dots=1+\frac{\alpha_{II}}{\gamma_{n}}(\frac{v_{0}^2-v_{n}^{2}}{2}),
\ee
Therefore, the exact expression for $x_{n}$, including all contributions, can be writen as 
\be
\l{KUn}
x_{n}=1+\frac{\alpha}{\gamma_{n}}(\frac{v_{0}^2-v_{n}^{2}}{2})+\sum_{s=1}^{s=n}\alpha^{s}{X'}_{n}^{(s)},
\ee
where ${X'}_{n}^{(s)}$ represents all the non-dominant contributions from $X_{n}^{(s)}$. Then Eq. (\ref{KUn}) can be rearranged as
\be
\l{Xlinha}
(\frac{v_{0}^2}{2}-\frac{K}{\alpha}\frac{1}{r_{0}})-(\frac{v_{n}^2}{2}-\frac{K}{\alpha}\frac{1}{r_{n}})=-\sum_{s=1}^{s=n}\alpha^{s-1}\gamma_{n}{X'}_{n}^{(s)}.
\ee
We notice the appearing of the effective potential
\be
\l{Un}
U(r)=-\frac{K}{\alpha}\frac{1}{r},
\ee
which is a conservative field in the measure that the non-dominant contributions ${X'}_{n}^{(s)}$ can be neglected. Energy is always conserved but the expression (\ref{Un}) for the effective potential is exact only in the limit of $n\rightarrow\infty,$ which would be the only way of justifying, in absolute terms, the neglecting of the right-hand-side of Eq. (\ref{Xlinha}). So, $U(r)$ is just an idealized useful limiting concept. We can see from Eq. (\ref{Un}), in the case of gravitational interaction, that, as expected
\be
\l{GM}
\frac{K}{\alpha}=GM,
\ee
where $G$ is the gravitational constant and $M$ is the central mass. 

From Eq. (\ref{Xn1}) we have that
\be
{X'}_{n}^{(1)}=\frac{n\gamma_{n}}{2}+{\cal{O}}(\alpha)=\frac{nK}{2r_{0}}+{\cal{O}}(\alpha),
\ee
and so the right-hand-side of Eq. (\ref{Xlinha}) can be expanded as 
\be
\l{Xlinha1}
-\sum_{s=1}^{s=n}\alpha^{s-1}\frac{K}{r_{0}}{X'}_{n}^{(s)}=-\frac{K^{2}n}{2r_{0}^{2}}
+{\cal{O}}(\alpha).
\ee
For a nearly radial orbit, $n$, in a first approximation, is  proportional to the traversed  space distance $\Delta r= r_{n}-r_{0}$, and so, it corresponds to the observed anomalous acceleration.  This is not so for a nearly circular orbit and so this anomalous acceleration cannot be observed on planetary ephemeris.
On the other hand, from Eqs. (\ref{vs},\ref{Vjn0},\ref{sxn}) 
\be
v_{n}\approx v_{0}-n\frac{K}{r_{0}}+{\cal{O}}(\alpha),
\ee
and  
\be
\l{sxn1}
x_{n}\approx1+\alpha{\Big \{}n(v_{0}-\frac{n\frac{K}{r_{0}}}{2}){\Big \}}+{\cal{O}}(\alpha^2),
\ee
so that
\be
\frac{dn}{dr_{n}}=\frac{1}{\alpha r_{0}v_{n}},
\ee
\be
a_{P}=\frac{K^2}{2\alpha_{I}r_{0}^{3}v_{n}}=\frac{a_{1}K}{2r_{0}v_{n}},
\ee
and
\be
\frac{da_{P}}{dr_{n}}=\frac{a^{2}_{1}K}{2r_{0}v^{3}_{n}}=\frac{a_{1}a_{P}}{v^{2}_{n}},
\ee

where $a_{P}$ is the extra acceleration and $a_{1}$ is the value of the standard acceleration at $r_{0}=1UA$. 
If we take \cite {gr-qc/0104064} $a_{P}\simeq(8,74\pm0.94)\times10^{-10}m/s^{2}$ we have
$$\Delta_{I}=2v_{n}\frac{a_{P}}{a_{1}}\approx v_{n}\times10^{-8,}$$
$$\alpha_{I}\approx{v}_{n}10^{-28}\quad[MKS]$$ and $$\frac{da_{P}}{dr_{n}}\approx\frac{1}{v^{2}_{n}}10^{-11}\quad[MKS]$$
For (astronomically) large distances $n$, and therefore $v_{n}$, changes very slowly so that taking $a_{P}$ as a constant is a good approximation in face of the present accuracy on the $a_{P}$ experimental determination.

\end{document}